\documentclass{iopart}
\usepackage{amssymb}
\usepackage{iopams}
\usepackage[dvips]{graphicx}

\catcode`\@=11
\renewcommand\footnoterule{%
  \kern-3\p@
  \hrule\@width.4\columnwidth
  \kern2.6\p@}
\renewcommand\@makefntext[1]{%
    \parindent 1em\noindent
    \hb@xt@1.8em{\hss$^{\@thefnmark}$)}\hspace{2pt}%
    \footnotesize\rmfamily#1}  
\def\@makefnmark{\hspace{.5pt}\hbox{$^{\@thefnmark}$%
\hspace{-1pt})}} 

\def\be{\begin{equation}}
\def\ee{\end{equation}}
\def\ben{\begin{displaymath}}
\def\een{\end{displaymath}}
\def\ba{\begin{array}{c}}
\def\bal{\begin{array}{l}}
\def\ea{\end{array}}
\def\p{\partial}

\newcommand{\nn}{\nonumber}

\def\RR{\mathbb{R}}

\begin{document}

\title[Dynamics of charged fluids and $1/\ell$
perturbation expansions]{Dynamics of charged fluids
and $1/\ell$
perturbation expansions}

\author{Miloslav Znojil}
\address{Nuclear Physics Institute ASCR,
250 68 \v{R}e\v{z}, Czech Republic}
\ead{znojil@ujf.cas.cz}

\author{Uwe G\"unther}

\address{Research Center Dresden-Rossendorf,  POB 510119,
D-01314 Dresden, Germany}
\ead{u.guenther@fzd.de}

\begin{abstract}
Some features of the calculation of fluid dynamo systems
(spherically symmetric $\alpha^2-$dynamos) in magnetohydrodynamics
are studied, the problem connected with the presence of mixed
(Robin) boundary conditions is addressed and a new treatment for it
is proposed. The perturbation formalism of large$-\ell$ expansions
is shown applicable and its main technical steps are outlined.
\end{abstract}

%

\section{Introduction
}

The magnetohydrodynamics of  conductive fluids plays an important
role in the explanation of the existence and stability of the
magnetic field of the Earth as well as of the magnetic fields of
stars and galaxies studied in astrophysics \cite{earth} -
\cite{flips}. In the theoretical description of the motion of such
fluids an important role is played by the induction of magnetic
fields which are able to generate a genuine global dynamo effect
 \cite{dynamoRiga}. For certain spherically symmetric
field configurations after a mean field approximation a so called
$\alpha^2-$dynamo model can be obtained \cite{whereverderived}.
Its description may be reduced \cite{whereverderived} to the
coupled pair of ordinary differential phenomenological equations
\cite{GS-jmp2003,jmpus}
 \be
 -\p ^2_r \phi(r) +V_u(r)\,\phi(r)- \alpha(r)\,\chi(r)
 = -\lambda\,\phi(r),
 \label{equationsa}
 \ee
 \be
 -\p ^2_r \chi(r) +V_d(r)\,\chi(r)
 +\p _r \alpha(r) \p _r \phi(r) -V_m(r)\,\phi(r)
 = -\lambda\,\chi(r)\,.
 \label{equationsb}
 \ee
They are defined on a finite interval of a single coordinate $r
\in (0,R)$ with $R=1$ after a re-scaling. For the purely kinematic
factors one has to set $V_u(r) = V_d(r) =
V_m(r)/\alpha(r)=\ell(\ell+1)/r^2$ where the integer parameter
$\ell = 0, 1, \ldots$ coincides with the angular mode number of
the field. The only input information about the flow of the
charged fluid (or plasma) is carried by the shape of the so called
$\alpha-$profile $\alpha(r)$.

Let us now formulate our present task as the construction of the
solutions of equations (\ref{equationsa}) and (\ref{equationsb})
specified by realistic physical boundary conditions
 \be
  \phi(0)=0, \ \ \ \ \left [\p _r \phi(r)\right |_{r=R}
 +\frac{\ell}{R}\,\phi(R) =0,
 \label{bca}
 \ee
 \be
  \chi(0)=\chi(R)=0\,.
 \label{bcb}
 \ee
The presentation of our results will start by an outline of
consequences of the difference between the latter two boundary
conditions in section \ref{druha}. In section \ref{druhabe} we
show how this difference diminishes in proportion to the quantity
$1/\ell$. This, of course, indicates the possible applicability of
a standard perturbation expansion in this parameter. After a few
introductory technical remarks made on such an approach in the
separate Appendix A we demonstrate, in the next section
\ref{thron}, that the parameter $1/\ell$ can really play the role
of a measure of perturbation of the solvable $\ell \to \infty$
limit. In the subsequent sections \ref{hufu} and  \ref{cfhufu} we
then describe the explicit realization of such a programme in more
detail. Finally, section \ref{suumi} summarizes all the key
ingredients of our present new method of solution of the
magnetohydrodynamical $\alpha^2-$dynamo problem by the asymptotic
series in terms of certain rational powers of $1/\ell$.

\section{The dynamo  problem in two-space formulation
 \label{druha} }

There exist several formal analogies of the dynamo problem with
quantum mechanics
\cite{sokoloff,bar-shu-1,meinel,meinel-brandenburg}. We intend to
employ here some of these analogies as a methodical guide. A deeper
discussion of this aspect of the problem can be found in
refs.~\cite{jmpus} - \cite{GS-cz3}. Nevertheless, one should keep in
mind that these analogies are violated, first of all, by the
presence of mixed-type boundary conditions.

\subsection{The doublet of bases \label{archit}}


The most challenging mathematical feature of the physical boundary
conditions (\ref{bca}) and (\ref{bcb})  is that they prescribe the
use of {\em different} spaces, say, ${\cal V}_u$ and ${\cal V}_d$
for the respective channel functions $\phi(r)$ (representing the
so called poloidal mode of the magnetic field) and $\chi(r)$
(representing the complementary, toroidal mode). In a compactified
Dirac's bra-ket notation this means that one is forced to employ
different symbols for elements of each of these spaces. Let us
employ the usual vector or ket-symbol $|\,\chi\rangle \in {\cal
V}_d$ in place of the function $\chi(r)$ emphasizing that for this
Hilbert-space element the Dirichlet boundary condition (\ref{bcb})
at $r=R=1$ is entirely standard. A modified abbreviation
$|\,\phi\} \in {\cal V}_u$ will be introduced for $\phi(r)$ which
obeys the less standard (mixed, Robin) boundary condition
(\ref{bca}) containing the derivative.

In the new notation one may introduce orthonormalized bases of
eigenvectors resulting from the spectral representations of
self-adjoint differential operators in any of the two vector
spaces ${\cal V}_u$ and ${\cal V}_d$. We shall postulate
 \be
 -\p ^2_r  +V_u(r) = \sum_{m=0}^\infty\,|\,m\}\,\tau_m\,\{ m|,
 \label{hornak}
  \ee
 \be
 -\p ^2_r  +V_d(r) = \sum_{n=0}^\infty\,
 |\,n\rangle\,\varrho_n\,\langle n|,
 \label{dolnak}
 \ee
knowing that due to the simplicity of $V_{u,d}(r)$ and due to the
coincidence of the boundary conditions in the origin, all the
basis functions $\phi_n(r) =\{r |\,n\}$ as well as $\chi_n(r)
=\langle r |\,n\rangle$ must be proportional to Bessel special
functions (ref. \cite{Oleg} may be consulted for more details).
The basis states have the same closed form with different scaling,
 \be \fl \ \ \ \ \ \ \ \ \ \
 \varphi_m(r)=M_mr^{1/2}J_{\ell+1/2}\left (
 \sqrt{\tau_m}\,r\right ), \ \ \ \
 \chi_n(r)=N_nr^{1/2}J_{\ell+1/2}\left (
 \sqrt{\varrho_n}\,r\right )\,.
 \label{solutionsg}
 \ee
The kinetic operator eigenvalues $\tau_p$ and $\varrho_q$ are
given as roots of the appropriate combinations of Bessel functions
at $r=R =1$. Their numerical values are available with arbitrary
precision, therefore.

More care must be paid to the operators which couple the channels.
Once they represent a map between ${\cal V}_u$ and ${\cal V}_d$ we
can only employ a non-diagonal, full-matrix formula for the
operator of the multiplication by the function $\alpha(r)$ in
eq.~(\ref{equationsa}),
 \be
 \alpha(r)=
 \sum_{k,j=0}^\infty\,|\,k\}\,\{k|\,\alpha(r)|\,j\rangle\,\langle j|\,.
 \label{preformer}
 \ee
The technique of the evaluation of the matrix elements $\{
k|\,\alpha(r) |\,j \rangle$ (by integration) is standard. {\it
Mutatis mutandis,} the same comment applies to the other
channel-coupling operator
 \be
 \p _r \alpha(r) \p _r -V_m(r)
 = - \sum_{k,j=0}^\infty\,|\,k\rangle\,\omega_{k,j}\,\{ j|
 \label{former}
 \ee
in
eq.~(\ref{equationsb}).

\subsection{Linear algebraic form of the coupled equations}

The spectral series (\ref{hornak}) and (\ref{dolnak}) as well as
the full-matrix formulae (\ref{preformer}) and (\ref{former})
should be inserted in the coupled set of differential equations
(\ref{equationsa}) and (\ref{equationsb}). Using the pair of the
natural ans\"{a}tze
 \be
 |\,\phi\} = \sum_{k=0}^\infty\,|\,k\}\, \phi_k, \ \ \ \ \
 |\,\chi\rangle = \sum_{j=0}^\infty\,|\,j\rangle\, \chi_j,
 \label{ansa}
 \ee
we get the set of relations
 \be
 \sum_{m=0}^\infty\,|\,m\}\,(\tau_m+\lambda)\,\phi_m=
 \sum_{k,j=0}^\infty\,|\,k\}\,\{k|\,\alpha(r)|\,j\rangle\,\chi_j\,,
 \label{equationsasajj}
 \ee
 \be
 \sum_{k,j=0}^\infty\,|\,k\rangle\,\omega_{k,j}\,\phi_j=
 \sum_{n=0}^\infty\,|\,n\rangle\,(\varrho_n+\lambda)\,\chi_n\,.
 \label{equationsbsbjj}
 \ee
Once we assume the completeness of both orthonormalized basis sets
$\{\,|\,p\}\,\}_{p=0}^\infty$ and $\{\,|\,q\rangle\, \}_{
q=0}^\infty$ spanning respective infinite-dimensional vector
spaces ${\cal V}_u$ and ${\cal V}_d$ we are allowed to multiply
the previous pair of equations by the respective bra-vectors
$\{p|$ and $\langle q|$ from the left. This leads to the final
formulation of the dynamo problem in the form of the infinite set
of linear algebraic equations,
 \be
 \left (\tau_p+\lambda \right )\,\phi_p=
 \sum_{j=0}^\infty\,\{p|\,\alpha(r)|\,j\rangle\,\chi_j\,,\ \ \ \ \
 p = 0, 1, \ldots\,,
 \label{equationsasa}
 \ee
 \be
 \sum_{j=0}^\infty\,\omega_{q,j}\,\phi_j=
 \left (\varrho_q+\lambda\right )\,\chi_q\,,\ \ \ \ \ \ \ \
 q = 0, 1, \ldots
 \,.
 \label{equationsbsb}
 \ee
These equations may only be solved numerically (in this case it
makes sense to start from the elementary elimination, say, of all
the unknown quantities $\phi_p$ using eq.~(\ref{equationsasa})) or
perturbatively (in the latter case it seems better to preserve the
linearity of the whole set in $\lambda$).

\subsection{The simplified model with constant
 $\alpha(r)$ \label{tretib} }

Both operators (\ref{preformer}) and (\ref{former}) become
perceivably simpler for constant $\alpha-$profiles. It allows us
to simplify eq.~(\ref{equationsasajj}),
 \ben
 \sum_{k=0}^\infty\,|\,k\}\,(\tau_k+\lambda)\,\phi_k
 =\alpha_0\,
 \sum_{k,j=0}^\infty\,|\,k\}\,\{k|\,j\rangle\,\chi_j\,.
 \een
In parallel, a factorization of the matrix
$\omega_{kj}$ in eq.~(\ref{former}),
 \ben
 \p _r \alpha_0 \p _r -\alpha_0\ell(\ell+1)/r^2
 = - \alpha_0 \sum_{k,j=0}^\infty\,|\,k\rangle\,\langle k|
 \, j\}\,\tau_j\,\{
 j|\,
 \een
converts eq.~(\ref{equationsbsbjj}) into the simpler relation
 \ben
 \alpha_0\,\sum_{n,m=0}^\infty\,|\,n\rangle\,\langle n
 |\,m\}\,\tau_m\,\phi_m=
 \sum_{n=0}^\infty\,|\,n\rangle\,(\varrho_n+\lambda)\,\chi_n\,.
 \een
As a result one gets the pair of coupled linear algebraic equations
\begin{eqnarray}
&& \left (\tau_k+\lambda \right )\,\phi_k
 =\alpha_0\,
 \sum_{j=0}^\infty\,\{k|\,j\rangle\,\chi_j\,,
 \label{matrixpra}
 \\
&& \alpha_0\,\sum_{m=0}^\infty\,\langle k |\,m\}\,\tau_m\,\phi_m=
 \left (\varrho_k+\lambda \right )\,\chi_k\,,\ \ \ \ \ \ k=0, 1, \ldots
 \label{matrixpr}\label{matrixprbe}
\end{eqnarray}
which replaces eqs.~(\ref{equationsasa}) and (\ref{equationsbsb})
at $\alpha(r)=\alpha_0$. For numerical purposes we may easily
eliminate one of these relations and get the single Feshbach-type
set for $k=0, 1, \ldots$,
 \be
 \sum_{m=0}^\infty\,
 \left [
 \left (\tau_k+\lambda \right )\,\delta_{k,m}
 -\alpha_0^2\,
 \sum_{j=0}^\infty\,\{k|\,j\rangle\,\frac{1}{
 \left (\varrho_j+\lambda \right
 )}\,\langle j |\,m\}\,\tau_m\,
 \right ]\,
 \phi_m=0
 \label{matrixpras}
 \ee
or, alternatively,
 \be
 \sum_{m=0}^\infty\,
 \left [
 \left (\varrho_k+\lambda \right )\,\delta_{k,m}
 -\alpha_0^2\,
 \sum_{j=0}^\infty\,\{k|\,j\rangle\,\frac{\tau_j}{
 \left (\tau_j+\lambda \right
 )}\,\langle j |\,m\}\,
 \right ]\,
 \chi_m=0\,,
 \label{matrixprbsc}
 \ee
i.e., after a slight modification,
 \ben
 \sum_{m=0}^\infty\,
 \left [
 \left (\varrho_k+\lambda -\alpha_0^2\right )\,\delta_{k,m}
 +\alpha_0^2\,
 \sum_{j=0}^\infty\,\{k|\,j\rangle\,\frac{\lambda}
 {\left (\tau_j+\lambda \right
 )}\,\langle j |\,m\}\,
 \right ]\,
 \chi_m=0\,.
 \een
We see that the knowledge of the single matrix of overlaps
$\{k|\,j\rangle$ is the only input needed for the standard
numerical solution of these equations. Unfortunately, the latter
observation is in fact of no immediate impact upon applications
since the choice of the constant $\alpha(r) =\alpha_0\neq 0$ is
not too realistic \cite{earth,whereverderived}. Even in a more
formal evaluation it seems oversimplified as it allows many
specific relations between its basis states (\ref{solutionsg}).
One should note that the eigenvalues even become available as
roots of a certain quadratic superposition of Bessel functions as
a consequence \cite{whereverderived}. In this sense our toy
problem $\alpha(r) =\alpha_0$ remains exactly solvable.
Nevertheless, it deserves full attention as a very useful
methodical guide to more realistic situations. Our present
ambition will concentrate upon the proposal and description of its
new, promising and fairly natural {\em perturbative} treatment.

\section{The confluence of bases in spaces
${\cal V}_u$ and ${\cal V}_d$ at
 $\ell \gg 1$ \label{druhabe}}

There exist several important differences between the present
eigenvalue problem~(\ref{equationsasa}) + (\ref{equationsbsb}) and
its quantum-mechanical coupled-channel analogues \cite{cc}.
Firstly, one should note the difference in the sign convention
(the energies of quantum mechanics would be $E=-\lambda$).
Secondly, the present eigenvalue problem is not self-adjoint in
the form which would be usual in quantum mechanics (consult again
ref.~\cite{Oleg} for a deeper discussion of this aspect). Thirdly,
one usually does not encounter Robin boundary conditions in
quantum mechanics. In this sense our present problem is
perceivably more complicated.

\subsection{Boundary conditions}

A formal key to the simplification of equations
(\ref{equationsasa}) + (\ref{equationsbsb}) can be sought in a
{\em decrease} of the difference between the Robin and Dirichlet
boundary conditions (\ref{bca}) and (\ref{bcb}) at $r=R=1$ with
the growth of $\ell$. In a way which significantly weakens the
above-mentioned constant-profile assumption $\alpha(r)=\alpha_0$
let us merely assume now that $\alpha(r)$ remains more or less
constant in a small vicinity of $R=1$. In this case with $ r
\approx R$, equations (\ref{equationsa}) and (\ref{equationsb})
acquire the following approximate form,
 \ben \fl
 -\p ^2_r \phi(r) +
 \left (\kappa^2+\lambda \right )\,\phi(r)- \alpha(R)\,\chi(r)
 =0,\ \ \ \ \ \
 \kappa^2=\frac{\ell(\ell+1)}{R^2}\,,
 \een
 \be \fl
 -\p ^2_r \chi(r) +\left (\kappa^2+\lambda \right )\,\chi(r)
 +\alpha(R) \,\p _r^2 \phi(r)
 -\alpha(R)\,\kappa^2\,\phi(r)
 = 0\,, \ \ \ \ r \approx R\,.
 \label{equationsap}
 \ee
This system (with constant coefficients) possesses
eigenfrequencies $\mu=\mu[\lambda,\kappa,\alpha(R)]$ obtainable
directly from the corresponding characteristic equation
 \ben
 \left (
 \mu^2+\kappa^2+\lambda
 \right )^2=\alpha^2(R)\,
 \left (
 \mu^2+\kappa^2
 \right )\,.
 \een
In the light of boundary conditions at $r=R$ we may expect that
$\chi(r)=D\,\sin \mu (r-R)$ and $\phi(r)=C\,\sin \mu (r-S)$ where
$S \approx R$ is unknown and where $R=1$. An estimate of the value
of $S$ may be deduced via insertion from eq.~(\ref{bca}),
 \be
  \mu\,\cos \mu (R-S)
 +\frac{\ell}{R}\,\sin \mu (R-S)
  =0\,.
  \label{ressu}
 \ee
For all $\ell > 0$ this enables us to conclude that $S>R=1$ so
that $\tau_k<\varrho_k$ at $k=0, 1, \ldots$.

At very large angular excitations the following explicit
frequency-independent estimate results from eq.~(\ref{ressu}),
 \be
 S = 1 +\frac{1}{\ell} +{\cal O}\left (\frac{1}{\ell}\right )^2\,.
  \label{ressumee}
 \ee
We see that the two boundary conditions (\ref{bca}) and
(\ref{bcb}) and, hence, also the two bases spanning the spaces
${\cal V}_u$ and ${\cal V}_d$ will coincide in the limit $\ell \to
\infty$,
 \be
 \lim_{\ell \to \infty} \tau_n =\varrho_n, \ \ \ \
 \lim_{\ell \to \infty} \,|\,n\} \to |\,n\rangle, \ \ \ \ \ n = 0, 1,
 \ldots\,.
 \ee
This is an important observation. From the practical point of view
it means that we shall have
 \be
 \lim_{\ell \to \infty} \langle q|\,n\} = \delta_{q,n}\,
 \ \ \ \ \ \
 \lim_{\ell \to \infty} \langle q|\,\alpha(r) |\,n\}
 = \langle q|\,\alpha(r) |\,n\rangle
 \label{limort}
 \ee
for all the indices $q\geq 0$ and $n\geq 0$ and for all reasonable
functions $\alpha(r)$.

\subsection{Illustration: Exact solvability of the $\ell \gg 1$ model at a
constant $\alpha$ \label{treti} }

As the only dynamical input in eqs.~(\ref{equationsa}) and
(\ref{equationsb}) the $\alpha-$profile function is a key to all
the practical and phenomenological considerations. At the same
time, all its not too large deviations from a constant mean
$\alpha-$profile   $\alpha_0$ may be treated perturbatively in the
way outlined in ref. \cite{Oleg}.

Let us now turn attention to the unperturbed problem with
$\alpha(r)=\alpha_0$ where we add as another assumption that $\ell
\gg 1$.  Due to the limit of the orthogonality rule (\ref{limort})
the resulting doubly simplified algebraic eqs.~(\ref{matrixpr})
become completely decoupled. At every index $k$ the condition of
vanishing secular determinant remains trivial,
 \ben
 \det\,
 \left (
 \begin{array}{cc}
 \tau_k+\lambda &-\alpha_0\\
 -\alpha_0\tau_k&\varrho_k+\lambda
 \ea
 \right )\approx 0\,, \ \ \ \ \ \tau_k\approx \varrho_k\,,
 \ \ \ \ \ \ \ \ell \gg 1\,
 \een
and its closed solution exists,
 \be
 \lambda_{1,2}=\frac{1}{2}
 \left [
 -\tau_k-\varrho_k\pm \sqrt{\left (\tau_k-\varrho_k
 \right )^2+4\varrho_k\alpha_0^2}
 \right ]\,.
 \ee
In the normalization $\phi_k=\alpha_0$ the closed form of the
coefficients is equally elementary,
 \be
 \left (\chi_k
 \right )_{1,2}=\tau_k+\lambda_{1,2}=\frac{1}{2}
 \left [
 \tau_k-\varrho_k\pm \sqrt{\left (\tau_k-\varrho_k
 \right )^2+4\varrho_k\alpha_0^2}
 \right ]\,.
 \ee
As long as we have $\tau_k\approx \varrho_k$, the latter recipe
reproduces exactly the definition $\lambda=\lambda_k^{\pm}=
-\varrho_k\pm \alpha_0\sqrt{\varrho_k}$ of the spectrum as obtained
earlier under an alternative assumption of the high conductivity
\cite{high-conductivityb,high-conductivityc} of the dynamo's fluid
(cf. eq.~(13) in ref.~\cite{Oleg}).

\section{Effective simplifications of boundary conditions \label{thron}}

Several complications outlined in section \ref{druha} disappear when
one returns to the zeroth-order $\ell \to \infty$ approximation in
eq.~(\ref{ressumee}) \cite{Oleg}. For this reason we intend to pay
attention to the next, first-order level of approximation in
$1/\ell$.

\subsection{Approximation using a pair of Dirichlet boundary
conditions \label{transi}}

We feel strongly motivated by the observation that
eq.~(\ref{ressumee}) in fact {\em prescribes} an ``effective"
replacement of the Robin (i.e., ``difficult") boundary condition
(\ref{bca}) by the Dirichlet (i.e., ``easy") boundary condition,
say,
 \be
  \phi(0)=0, \ \ \ \ \phi(S) =0,\ \ \ \
 S = S(\ell)=1 +\frac{1}{\ell}\,.
 \label{bceff}
 \ee
One should note that the new condition is specified just by a
rightward shift of the end of the interval. An important
``user-friendly" feature of this shift should be seen in the fact
that on the first-order level of precision the value of $S(\ell)$
remains {\em independent} of all the other parameters. We may
expect that the approximation (\ref{bca}) $\to$ (\ref{bceff}) will
represent a fairly reliable and nontrivial approximation at all
angular mode numbers which are not too small. The idea might even
remain applicable in the very ``realistic" domain of the smallest
angular mode number $\ell$ where a further improvement of the
choice of $S(\ell)$ could be sought via eq.~(\ref{ressu}) whenever
necessary.

As next step a (purely formal) transition to all the real axis of
$r\in \RR$ will be mediated by the immersion of both coupled (i.e.,
poloidal and toroidal) subsystems of the whole system in infinitely
deep square wells $V^{(SQW)}_{Y}(r)$. They are defined as very large
(or infinite) everywhere except the interval $(0,Y)$ where they
should vanish,
 \be
 V^{(SQW)}_{Y}(r) = \left \{
 \begin{array}{ll}
  +\infty, \ \ \ & r \in (-\infty,0),\\
 0,\ \ \ & r \in (0,Y),\\
  +\infty, \ \ \ & r \in (Y,\infty).
 \ea
 \right .
 \label{sqwdef}
 \ee
This means that we change the definition of the kinematic factors
in eqs.~(\ref{equationsa}) and (\ref{equationsb}) where we set
$Y=S=1+1/\ell > 1$ and $Y=R=1$, respectively,
\begin{eqnarray}
 V_u(r) &=& V_u^{(SQW)}(r) =\frac{\ell(\ell+1)}{r^2}
  +V^{(SQW)}_{S}(r), \nn\\
 V_d(r)&=& V_d^{(SQW)}(r) =\frac{\ell(\ell+1)}{r^2}
  +V^{(SQW)}_{R}(r)\,.
 \label{redea}
\end{eqnarray}
The latter convention is purely formal and its use merely
emphasizes the essence of the introduction of the approximation
(\ref{bceff}).

\subsection{Models with smeared boundaries}

In the context of paragraph \ref{transi} the use of the spectral
series (\ref{dolnak}) remains based on closed solutions of the
Sturm-Liouville problem,
 \be
 -\p ^2_r \chi_n(r) +\frac{\ell(\ell+1)}{r^2}\,\chi_n(r)
 +V^{(SQW)}_{R}(r)\,\chi_n(r)
 = \varrho_n\,\chi(r)\,.
 \label{zerequationsa}
 \ee
The trick extends to the Robin boundary condition immediately. The
parallel construction  of the second auxiliary spectral series
(\ref{hornak}) is based on the solution of the similar equation
 \be
 -\p ^2_r \phi_n(r) +\frac{\ell(\ell+1)}{r^2}\,\phi_n(r)
 +V^{(SQW)}_{S}(r)\,\phi_n(r)
 = \tau_n\,\phi(r)\,.
 \label{zerequationsb}
 \ee
As long as the effect of the Dirichlet boundary conditions is
mimicked by the action of an infinitely deep square-well
potential, a perceivable reduction of the complexity of the basis
$|n\}$ {as well as} of the modified space ${\cal V}_{u}$ is
achieved. It is also more easy to evaluate quantities $\tau_n$.

In summary, the effective simplification of boundary conditions of
paragraph \ref{transi} looks satisfactory. It is possible to argue
that although the numerical {\em performance} of the resulting
simplified bases could be enhanced significantly by the
modification of boundary conditions, the {\em precision} of
results may still be kept under control by making the modification
``infinitesimal" (i.e., controllably small). One can only feel
dissatisfied by the non-analyticity of square wells (\ref{sqwdef})
which does not seem to open any easy way towards understanding of
the observable simplifications at $\ell \to \infty$.

For this reason it would make sense to replace the non-analytic
auxiliary square wells in eqs.~(\ref{zerequationsa}) and
(\ref{zerequationsb}) by some ``infinitesimally" modified {\em
analytic} approximants. One should add that the latter idea need
not necessarily be in any conflict even with the underlying
experimental setup. Moreover, a transition from the {\em
discontinuous} functions $V^{(SQW)}_{S,R}(r)$ to many of their
available respective analytic alternatives $U^{(AA)}_{u,d}(r)$ may
be expected better tuned to the {\em analytic} essence of
perturbation methods.

In this spirit let us introduce models with smeared boundaries
characterized  by a final redefinition of the corresponding
non-analytic functions (\ref{redea}),
\begin{eqnarray}
 V_u(r) &=&V_u^{(AA)}(r) = \frac{\ell(\ell+1)}{r^2}
  +U^{(AA)}_{u}(r), \nn\\
 V_d(r) &=&V_d^{(AA)}(r)= \frac{\ell(\ell+1)}{r^2}
  +U^{(AA)}_{d}(r)\,.
 \label{redeinta}
\end{eqnarray}
The two requirements of a sufficiently precise fit to the square
wells,
 \be
 U^{(AA)}_{d}(r)\approx V^{(SQW)}_{R}(r)\,,\ \ \ \ \
 U^{(AA)}_{u}(r)\approx V^{(SQW)}_{S}(r)
 \label{blouma}
 \ee
are the only limitations of our free choice of the new {\em
analytic} auxiliary potentials~$U$.

\section{The idea of $1/\ell$ perturbation expansions   \label{hufu} }

The effective shift of one of the boundary conditions in
eq.~(\ref{bceff}) {obviously} {\em introduces} a ``hidden" {\em
kinematical} parameter $1/\ell$ in the {\em dynamics} of our
present $\alpha^2-$dynamo models. The numerical smallness of this
parameter is one of our present most important observations. We
believe that the existence of such a small parameter offers  an
entirely new hint and encouragement for a new development and/or
(typically, perturbative) simplifications of many existing
practical calculations.

In principle, a prospective way of doing so might consist, e.g.,
in a systematic improvement of the linear-algebraic and
weighted-residual approximation techniques based on the truncation
of the infinite matrices to their finite-size approximants. A
complementary picture might be provided by the Fourier-series
based Galerkin techniques of ref.~\cite{Oleg} etc. Still, we
believe that the best use of the smallness of $1/\ell$ in our
present $\alpha^2$--dynamo models can be achieved via a direct use
of the various large$-\ell$ expansions as tested, in the various
contexts, e.g., in refs. \cite{Bjerrum} -- \cite{Bjerrumi}.

Unfortunately, the actual potential of the large$-\ell$ expansions
has only rarely been tested out of its natural quantum-mechanical
domain. Thus, only this experience as outlined briefly in Appendix
A below is, at present, available as our preliminary methodical
guide. Still, we believe that it offers a sufficiently strong
encouragement of the transfer of the $1/\ell$ asymptotic-series
techniques to the present MHD eqs.~(\ref{equationsa}) and
(\ref{equationsb}) accompanied by the boundary conditions
(\ref{bca}) and (\ref{bcb}).

Let us now complement the outline of the method as presented in
Appendix A by a more detailed account of its features which
transcend the routine applications, say, of
refs.~\cite{Fluegge,onenpan}.

\subsection{Coupled channels
and the smeared boundary conditions
 }

In the generic context of $1/\ell-$expansions we have to clarify,
first of all, the explicit perturbation account of the coupling of
the poloidal and toroidal modes. In the sequel we shall pick up
again equations (\ref{equationsasa}) and (\ref{equationsbsb}) with
$\alpha(r)=\alpha_0$. Formally, this enables us to re-arrange the
original differential equations,
 \be
 -\p ^2_r \phi(r) +V_u(r)\,\phi(r)- \alpha_0\,\chi(r)
 = -\lambda\,\phi(r),
 \label{equationsaww}
 \ee
 \be
 -\p ^2_r \chi(r) +V_d(r)\,\chi(r)
 -
 \alpha_0\,\left [ \alpha_0\,\chi(r)
 -\lambda\,\phi(r)
 \right ]
 = -\lambda\,\chi(r)\,.
 \label{equationsbww}
 \ee
This simplification will also enhance the transparency of our
considerations, paving the way towards their extension to more
realistic nonconstant $\alpha-$profiles.

In the $1/\ell-$series perspective of Appendix A, we have to add now
a discussion of a few specific  features of $\alpha^2$ dynamos. In a
preparatory step let us emphasize that the necessary underlying
transition to the smeared boundary conditions will imply the
necessity of the use of the coordinates along the whole real axis,
$r \in \RR$. For {any} smeared-boundary analytic potential
(\ref{redeinta}) we will then be allowed to write the corresponding
coupled eqs.~(\ref{equationsaww}) and (\ref{equationsbww}) in the
partitioned operator form
 \be
 \fl \ \ \ \ \ \ \ \ \ \
 \left (
 \begin{array}{cc}
 -\p ^2_r  +V_u(r)&-\alpha_0\\
 \alpha_0\lambda &-\p ^2_r +V_d(r)-\alpha_0^2
 \ea
 \right )\,
 \left (
 \ba
 \phi(r)\\
 \chi(r)
 \ea
 \right ) =-\lambda\,
 \left (
 \ba
 \phi(r)\\
 \chi(r)
 \ea
 \right )\,.
 \label{opera}
 \ee
As a consequence, we need not distinguish between the two distinct
linear spaces ${\cal V}_u$ and ${\cal V}_d$. We may search for the
smeared-boundary solutions corresponding to the low-lying
eigenvalues $-\lambda$ in the single and standard Hilbert space
$L_2(\RR)$. Nevertheless, in the perturbation context we shall
still be forced to employ the two different bases.

\subsection{Non-coincidence of the two local minima.}

As first step, paralleling the procedure of Appendix A, we make an
appropriate choice of the two boundary-simulating analytic
functions $U_{u,d}^{(AA)}(r)=U_{u,d}(r)$ of $r$ and expand
 \be
 \fl \ \ \ \ \ \
 V_{u,d}(r)=
 \frac{\ell(\ell+1)}{r^2} +U_{u,d}(r)= V_{u,d}({T}_{u,d})
 + V'_{u,d}({T_{u,d}})\,(r-{T}_{u,d})+
 \ldots\,.
 \label{tayl}
 \ee
Then the two parallel generalizations of eq.~(\ref{longrigh}),
viz.,
 \ben
  V'_{u}({T_{u}})=0\,, \ \ \ \ \
   V'_{d}({T_{d}})=0\,,
   \een
i.e., the two elementary quadratic equations
 \be
 {2\ell(\ell+1)}={T_{u}^3}\,U'_{u}(T_{u})={T_{d}^3}\,U'_{d}(T_{d})\,
 \label{fluvium}
   \ee
define in principle the pair of inverse functions
$T_{u,d}=T_{u,d}(\ell)$. By assumption they should grow with
$\ell$, say, in such a way that the two new independent measures
of smallness $1/T_{u,d}(\ell)$ converge to zero in the limit $\ell
\to \infty$.

The pair of the coordinates $T_{u,d}$ of the respective minima of
$V_{u,d}(r)$ will be different in general. In a way paralleling
the guidance by Appendix A the respective shapes of $V_{u,d}(r)$
near their minima will be specified by the next Taylor-series term
with the coefficient proportional to the minus-fourth power of the
scaling factor,
 \be \fl \ \ \ \ \
 V''_{u,d}(T_{u,d})=
 \frac{6\ell(\ell+1)}{T_{u,d}^4} +U''_{u,d}(T_{u,d}) =
 \frac{3\,U'_{u,d}(T_{u,d})}{T_{u,d}} +U''_{u,d}(T_{u,d})
 \
 \equiv\ 2\,\sigma_{u,d}^{-4}\,.
 \label{taylll}
 \ee
In this way we arrive at the zeroth-order descendant of
eq.~(\ref{opera}),
 \begin{eqnarray}
 \fl 
 \left (
 \begin{array}{cc}
 -\p ^2_\xi  +\xi^2+v_u&-\alpha_0\,\sigma_{u}^2\\
 \alpha_0\lambda^{[0]}\,\sigma_{d}^2 &-\p ^2_\zeta +\zeta^2+v_d-\alpha_0^2\,\sigma_{d}^2
 \ea
 \right )\,
 \left (
 \ba
 \phi^{[0]}(r)\\
 \chi^{[0]}(r)
 \ea
 \right )
 =-\lambda^{[0]}\,
 \left (
 \ba
 \sigma_{u}^2\,\phi^{[0]}(r)\\
 \sigma_{d}^2\,\chi^{[0]}(r)
 \ea
 \right )\,
 \label{operafre}
\end{eqnarray}
where we have introduced the additional abbreviation
$\sigma_{u,d}^2V_{u,d}(T_{u,d})\equiv v_{u,d}$ and where we have
to keep in mind that the symbol $r$ is just an abbreviation for
$r=T_{u}+\sigma_{u}\xi$ or $r=T_{d}+\sigma_{d}\zeta$.

\section{Feasibility of perturbation constructions \label{cfhufu} }

\subsection{Algebraic form of equations in zeroth-order limit
$\ell \to \infty$.}

As next step, guided by the notation used in subsection
\ref{archit}, we may innovate the spectral series (\ref{hornak}) and
(\ref{dolnak}) writing
 \be
 -\p ^2_\xi  +\xi^2
 = \sum_{m=0}^\infty\,|\,m\}\!\}\,\hat{\tau}_m\,\{\!\{ m|,
 \ \ \ \ \ \ \ \ \hat{\tau}_m=2m+1,
 \label{nehornak}
  \ee
 \be
 -\p ^2_\zeta +\zeta^2 = \sum_{n=0}^\infty\,
 |\,n\rangle\!\rangle \,\hat{\varrho}_n\,\langle\!\langle n|,
 \ \ \ \ \ \ \ \ \hat{\varrho}_n=2n+1
 \,.
 \label{nedolnak}
 \ee
The specific doubling of the bra-ket symbols has been chosen here to
underline the specific feature of both sets of eigenvectors: In the
zeroth-order perturbation limit $\ell \to \infty$ the eigenvalues
become well known at all subscripts, $\hat{\tau}_k=\hat{\varrho}_k=
2k+1$, and also the eigenvectors coincide with the two distinct
special cases of the harmonic-oscillator bases defined in terms of
Hermite polynomials \cite{Fluegge}.

It is necessary to add that even in the limit $\ell \to \infty$
the latter two harmonic-oscillator basis sets remain distinct in
general. A return to Appendix A reveals that due to the difference
between the Robin and Dirichlet boundary conditions at $\ell <
\infty$ the two choices of $V_{u,d}^{(AA)}$ [say, in the
anharmonic form sampled by eq.~(\ref{equaww})] must be necessarily
non-identical. Hence, also the resulting harmonic-oscillator basis
functions will differ in both of their $\ell-$dependent shifts
$T=T_{u,d}(\ell)$ and scaling factors $\sigma_{u,d}$. This means
that we have to re-write our unperturbed eq.~(\ref{operafre}) in
the spectral-series-like form paralleling eqs.~(\ref{preformer})
and (\ref{former}),
 \ben
 \fl \ \
 \left (
 \begin{array}{cc}
 \sum_{m=0}^\infty\,|\,m\}\!\}\,(\hat{\tau}_m+v_u)\,\{\!\{ m|
 &-\sigma_{u}^2\,\alpha_0\,
 \sum_{m,j=0}^\infty\,|\,m\}\!\} \{\!\{ m|\,j\rangle\!\rangle
  \langle\!\langle j|
 \\
 \sigma_{d}^2\,\alpha_0\lambda
 \,
 \sum_{n,p=0}^\infty\,
  |\,n\rangle\!\rangle \langle\!\langle n|
  \,p\}\!\} \{\!\{ p|
  &\sum_{n=0}^\infty\,
 |\,n\rangle\!\rangle
 \left (
 \hat{\varrho}_n+v_d-\sigma_{d}^2\alpha_0^2
 \right ) \langle\!\langle n|
 \ea
 \right )\,
 \left (
 \ba
 \phi(r)\\
 \chi(r)
 \ea
 \right )
 \een
 \be
 \ \ \ \ \ \ \
 =-\lambda\,
 \left (
 \ba
 \sigma_{u}^2\,\phi(r)\\
 \sigma_{d}^2\,\chi(r)
 \ea
 \right )\,, \ \ \ \ \ \ \ \ \ \ \ \ell \leq \infty\,.
 \label{bepera}
 \ee
Differently said, the resulting matrix problem remains purely
numerical even after the insertion of the present counterpart
 \be
 |\,\phi\}\!\} = \sum_{k=0}^\infty\,|\,k\}\!\}\, \phi_k, \ \ \ \ \
 |\,\chi\rangle\!\rangle = \sum_{j=0}^\infty\,|\,j\rangle\!\rangle
 \, \chi_j
 \label{beansa}
 \ee
of eqs.~(\ref{ansa}), giving the final infinite coupled set of
unperturbed equations
 \be
  \left (\frac{\hat{\tau}_k+v_u}{\sigma_{u}^2}+\lambda\, \right )\,\phi_k
 -\alpha_0\,
 \sum_{j=0}^\infty\,\{\!\{k|\,j\rangle\!\rangle\,\chi_j=0\,,
 \label{bematrixpra}
 \ee
 \be \fl \ \ \ \
    \left (\frac{\hat{\varrho}_k+v_d}{\sigma_{d}^2}+\lambda-\alpha_0^2
  \right )\,\chi_k+
 \alpha_0\,\lambda\,\sum_{m=0}^\infty\,\langle\!\langle k
 |\,m\}\!\}\,\phi_m=0\,,\ \
 \ \ \ \ k=0, 1, \ldots
 \label{bematrixpr}
 \label{bematrixprbe}
 \ee
which is just a smoothed-boundary parallel to
eqs.~(\ref{matrixpra}) and (\ref{matrixprbe}).

\subsection{Conditions of solvability of the zeroth-order equations
\label{presbuf} }

For a given pair of smoothed square wells $U^{(AA)}_{u,d}(r)$
which mimic boundary conditions the core of the applicability of
the large$-\ell$ perturbation theory may now be identified with a
guarantee of coincidence of the two bases in the limit $\ell \to
\infty$. Indeed, only in such a case one can find the application
of the standard textbook algorithms of perturbation expansions
\cite{onenpan} sufficiently economical and well motivated,
especially in comparison with the fairly efficient purely
numerical methods of linear algebra.

More explicitly, we require the exact solvability of the
zeroth-order version of the set of eqs.~(\ref{bematrixpra}) and
(\ref{bematrixprbe}). It may only be achieved when $|\,j\}\!\} \to
|\,j\rangle\!\rangle$ for all $j=0,1,\ldots$  in the limit $\ell \to
\infty$. In such a case we shall get the orthogonality rule
$\{\!\{k|\,j\rangle\!\rangle = \delta_{k,j}$ so that our set of
equations (\ref{bematrixpra}) and (\ref{bematrixprbe})  {\em
decouples} in infinitely many pairs of equations
 \ben \fl \ \ \ \
 \left (\frac{\hat{\tau}_k+v_u}{\sigma_{u}^2}+\lambda \right )\,\phi_k
 -\alpha_0\,\chi_k=0\,,\ \ \ \ \ \ \ \
 \alpha_0\,\lambda\,\phi_k+
 \left (\frac{\hat{\varrho}_k+v_d}{\sigma_{d}^2}+\lambda-\alpha_0^2 \right
 )\,\chi_k=0\,,
 \een
numbered by index $k=0, 1, \ldots$. With their secular determinant
vanishing,
 \ben
 \det\,
 \left (
 \begin{array}{cc}
 \hat{\tau}_k+v_u+\lambda\,\sigma_{u}^2 &-\alpha_0\,\sigma_{u}^2\\
 \alpha_0\,\lambda\,\sigma_{d}^2&\hat{\varrho}_k+v_d+
 \left (\lambda-\alpha_0^2\right )\,\sigma_{d}^2
 \ea
 \right )= 0\,,
 \ \ \ \ \ \ \ \ell \gg 1\,
 \een
we arrive at the exact solution
 \be \fl \ \ \ \ \
 \lambda_{1,2}=\frac{1}{2}
 \left [
 -\frac{\hat{\tau}_k+v_u}{\sigma_{u}^2}
 -\frac{\hat{\varrho}_k+v_d}{\sigma_{d}^2}\pm \sqrt{\left (
 \frac{\hat{\tau}_k+v_u}{\sigma_{u}^2}-\frac{\hat{\varrho}_k+v_d}{\sigma_{d}^2}
 \right )^2+4\,\frac{\hat{\tau}_k+v_u}{\sigma_{u}^2}
 \,{\alpha_0^2}{}}
 \right ]\,
 \label{letter}
 \ee
where we have to insert $\hat{\tau}_k= 2k+1=\hat{\varrho}_k$ and
$\sigma_{u}^2=\sigma_{d}^2$.

Although formulae (\ref{letter}) look formally very similar to the
examples we studied in subsection \ref{treti}, the key difference
lies in the fact that the coincidence of the bases may now be
achieved easily. In the light of eq.~(\ref{chov}) the necessary
and sufficient conditions of this coincidence $|\,j\}\!\} =
|\,j\rangle\!\rangle$ reads
 \be
 T_{u} =  T_{d} =T\,,\ \ \ \ \
  V''_{u}(T_{u})= V''_{d}(T_{d})\,.
  \label{husy}
  \ee
Due to eq.~(\ref{fluvium}) the first rule means that
 \be
 U'_{u}(T)= U'_{d}(T)\,
 \label{oskula}
 \ee
while the validity of eq.~(\ref{taylll}) then implies that the
rest of eq.~(\ref{husy}) is equivalent to the condition
 \be
 U''_{u}(T)= U''_{d}(T)\,.
 \label{oskulbe}
 \ee
We may conclude that the pair of eqs.~(\ref{oskula}) and
(\ref{oskulbe})  (i.e., in a geometric language the so called
osculation of the two curves) represents the necessary and
sufficient condition of the required coincidence $|\,j\}\!\} =
|\,j\rangle\!\rangle$ of the zeroth-order bases.

In practice we may expect that once we fix the value of $T\gg 1$, we
may guarantee the solvability of our present illustrative
zeroth-order $\alpha^2-$dynamo eigenvalue problem simply by the
choice of the pair of the smooth-boundary-simulating potentials
which obey the formula
 \be \fl \ \ \ \ \ \ \ \ \
 U_{u,d}(r)=U_{u,d}(T)+A\,(r-{T})+
 B\,(r-{T})^2+
 f_{u,d}(r-{T})\,(r-{T})^3\,.
 \label{taytr}
 \ee
This formula contains the {\em same} pair of parameters $A$ and
$B$ and {\em two different} functions $f_{u,v}(r-{T})$ which
remain regular at $r=T$ as well as {\em two different ``offsets"}
$U_{u,d}(T)$. Obviously, the latter formula offers enough freedom
for an arbitrarily precise and explicit necessary fit
(\ref{blouma}) of the boundary conditions.

\section{Summary \label{suumi} }

Despite our explicit knowledge of the  exact basis states
(\ref{solutionsg}), the immediate linear-algebraic recipe
(\ref{equationsasa}) + (\ref{equationsbsb}) for the construction
of the solutions of differential eqs.~(\ref{equationsa}) and
(\ref{equationsb}) is a numerical task. The necessary matrix
elements must be computed as integrals of the products of the
pairs of Bessel functions with given and, in principle, arbitrary
phenomenological input function $\alpha(r)$. The matrix elements
rarely remain available in closed form and sophisticated numerical
methods are necessary for their evaluation. Moreover, even if the
precision of the matrix elements themselves proves satisfactory
for a given $\alpha(r)$ in (\ref{preformer}) and (\ref{former}),
the final solution of the eigenvalue problem requires an
infinite-dimensional matrix diagonalization.

In this context we revealed that the quantity $1/\ell$ represents
a ``hidden" natural small parameter in the problem. This
encouraged us to search for an efficient non-numerical
construction of the solutions. An ambitious candidate has been
sought and found in the perturbation method of the so called
large$-\ell$ expansions. A detailed adaptation of the key
ingredients of this technique to the specific needs of the coupled
differential $\alpha^2-$dynamo equations has been performed here
in some detail.

As a first step of the realization of such a project the purely
Dirichlet specification of the toroidal mode $\chi(r)$ has been
selected and interpreted as if resulting from the action of an
infinitely deep square well $V^{(SQW)}(r)$ which vanishes
precisely inside the corresponding finite interval of $r$. This
trick gives an equivalent picture and at the same time it offers a
guide to the simplification of the treatment of the more
complicated poloidal mode $\phi(r)$ constrained by the Robin
boundary condition. In this sense the Robin constraint has been
re-interpreted as the same Dirichlet boundary condition modified
by its shift to some slightly higher value of an ``effective"
boundary point.

In the second step towards the realization of perturbation
solutions a ``softening"  of the ``rigid" boundary conditions has
been introduced via a replacement of the ``deep well"
$V^{(SQW)}(r)$ by an element of a large family of its suitable
analytic analogues and descendants $V^{(AA)}(r)$. For the sake of
definiteness the effects and consequences of this ``softening"
have been illustrated on the power-law well $V^{(AA)}(r)\sim \sum
a_K\,x^K$ with a suitable, not necessarily integer exponent $K \gg
1$. We reminded the readers how the $1/\ell$ technique works in
this older and, by the way, numerically extremely successful
\cite{Bjerrumd} application.

In the last third of our paper we demonstrated that and how the
``softening" of the boundary conditions opens a mathematically
consistent path towards the use of the inverse mode number $\ell$
as {\em the} effective small parameter of the theory of
spherically symmetric $\alpha^2-$dynamos. In particular we showed
that the asymmetric character of the coupling of the toroidal and
poloidal channels leads to certain unexpected and nontrivial
formal challenges and we were successful in finding their
resolution in the use of a certain flexibility in the choice of
the softened square wells.

Let us add in conclusion that the details and technical aspects of
our present method of construction will strongly depend on
assumptions concerning the structure of the input $\alpha-$profile
$\alpha(r)$. In fact, a strong sensitivity of the results on such
details has been observed and studied using a more standard
perturbation theory in ref.~\cite{Oleg}. Our present text skipped
the details of all the wealth of phenomena related to the
variations of $\alpha(r)$. In the light of our present emphasis on
the description and clarification of new methods we often
restricted our attention to the mere constant choice of
$\alpha(r)=\alpha_0$. This limitation seems to have left a lot of
space for its phenomenologically motivated weakening or even
complete removal.

\section*{Acknowledgements}

UG thanks O. Kirillov for useful and inspiring discussions and
appreciates the German Research Foundation DFG for the support by
the grant GE 682/12-3. MZ acknowledges the partial support by the
Saxonian Ministry of Science (grant Nr. 4-7531.50-04-844-06/5), by
GA\v{C}R (grant Nr. 202/07/1307) and by M\v{S}MT (``Doppler
Institute" project Nr. LC06002).

\newpage

\section*{Appendix A:  $1/\ell$ expansions in Quantum Mechanics
  }

Quantum anharmonic oscillators and their Schr\"{o}dinger equations
 \be
 \left [-\p ^2_r +V(r)\right ]
 \,\psi(r)
 = \varrho\,\psi(r)\,, \ \ \ \ \ \ \ V(r)=\frac{\ell(\ell+1)}{r^2}
  +\omega^2\,r^{2K}
 \label{equaww}
 \ee
are often studied in the domain of $\ell \gg 1$. Reviews
\cite{Bjerrumd} and \cite{Bjerrume} of the corresponding
techniques may be consulted for many technical details. In a
sketchy outline of these techniques let us emphasize that
eq.~(\ref{equaww}) might be also reinterpreted, in our present
$\alpha^2-$dynamo context, as one of the eligible smooth-boundary
alternatives to eq.~(\ref{zerequationsa}) or
(\ref{zerequationsb}). The single-channel Sturm-Liouville problem
(\ref{equaww}) defines then the basis states $|m\rangle$ or $|n\}$
entering the spectral series (\ref{dolnak}) or (\ref{hornak}),
respectively.

In a preparatory step of the current $1/\ell$ recipe let us
recollect that our function $V(r)$ may be re-written as the Taylor
series
 \be
 \fl \ \ \ \ \ \
 V(r)=
 \frac{\ell(\ell+1)}{r^2} +\omega^2\,r^K= V({T}) + V'({T})\,(r-{T})+
 \frac{1}{2}\, V''({T})\,(r-{T})^2+\ldots\,.
 \label{tay}
 \ee
At a point $T$ defined as the unique global minimum of $V(r)$ we
have, in particular,
 \be \fl \ \ \ \ \ \ \
 V'({T})=0 \ \ \ \Longrightarrow \ \ \
 \frac{2\ell(\ell+1)}{{T}^3} =K\,\omega^2\,{T}^{K-1}
 \ \ \ \Longrightarrow \ \ \ \,{T}^{k+2}=\frac{2\ell(\ell+1)}{K\,\omega^2}
 \,.
 \label{longrigh}
 \ee
This means that for large $\ell$ the coordinate ${T}={T}(\ell)$ is
also large. We may use the new measure of smallness $1/T$ in place
of the original small parameter $1/\ell$ in the formulae
 \ben
 V({T})=
 \frac{K\omega^2{T}^{K+2}}{2{T}^2}
 +\omega^2\,{T}^K=\omega^2\,{T}^K\,(1+K/2)\,,\ \ \ \ V'{({T})}=0\,,
 \een
 \ben
 \fl
 \ \ \ \
 V''({T})=
 {K\,\omega^2{T}^{K-2}}\,(K+2)\,,\ \ \ \ \
 V'''({T})=
 {K\,\omega^2{T}^{K-3}}\,(K^2-3K-10)\,,\ \ \ldots\,.
  \een
We insert them in eqs.~(\ref{tay}) and (\ref{equaww}) and change
variables in such a way that
 \be
 r-{T}=\sigma\,\xi, \ \ \ \
 \sigma^4=\frac{2}{
 K(K+2)\omega^2{T}^{K-2}
 }\,.
 \label{chov}
 \ee
This reduces eq.~(\ref{equaww}) to the formally equivalent
equation
 \ben
 -\p ^2_r \psi(r)
 +\frac{K(K+2)}{2}\,{\omega^2{T}^{K-2}}\,(r-{T})^2
 \,\psi(r)+
 \een
 \be \fl \ \ \ \ \ \ \ \ \ \ \
 +\frac{K(K^2-3K-10)}{6}\,{\omega^2{T}^{K-3}}\,(r-{T})^3
 \,\psi(r)+\ldots
 = [\varrho-V({T})]\,\psi(r)\,
 \label{eqww}
 \ee
and, after rescaling, to the perturbed harmonic oscillator,
 \be
 \left [
 -\p ^2_\xi
 +\xi^2
 +
 \kappa_3\,\xi^3+
 \kappa_4\,\xi^4
 +\ldots
 \right ]\,\psi({T}+\sigma\,\xi)
 = \varepsilon\,\psi({T}+\sigma\,\xi)\,.
 \label{erw}
 \ee
In the zeroth-order approximation its low-lying unperturbed
spectrum is well known and equidistant,
 \ben
 \varepsilon=\sigma^2\,[\varrho
 -V({T})]\approx \varepsilon_0= 1, 3, 5, \ldots\,
 \een
and all the higher-order perturbation corrections may be evaluated
easily \cite{Fluegge}. Moreover, their size may be made
arbitrarily small by the choice of a sufficiently large $\ell$
since it is easy to show that we have $\kappa_3 = {\cal O}\left
({T}^{-1/2-K/4}\right )$, $\kappa_4 = {\cal O}\left
({T}^{-1-K/2}\right )$ etc. This means that we get the pure and
exactly solvable harmonic oscillator in the limit $\ell \to
\infty$. Perturbation theory helps us then to evaluate the
corrections at any finite $\ell<\infty$ and for all the low-lying
levels with $n=0, 1, \ldots$,
 \be
 \varrho_n = V(T) +\sigma^{-2}\,(\varepsilon_0 +
 \varepsilon_1+\ldots)
 \label{perser}
 \ee
with $\varepsilon_j={\cal O}(\kappa_{j+2})= {\cal O}\left
({T}^{-j/2-jK/4}\right )$ for all perturbation orders
$j=1,2,\ldots$, i.e., with
 \ben
 \varrho_n =\omega^2(1+K/2) \,T^K
 + \omega\,(2n+1)\,\sqrt{K(1+K/2)}\,T^{K/2-1} + \ldots
 \een
in our illustrative example.


Marginally, let us note that the fact of the large size of the shift
of the position of the global minimum $T \gg 1$ of the complete, so
called ``effective" potential term $V_{eff}(r)=\ell(\ell+1)/r^2 +
V(r)$ in the Schr\"{o}dinger equation resolves, as a byproduct, also
the well known puzzle that our {\em approximative} wave functions
are, in general, allowed to become singular near the pole of the
centrifugal component of $V_{eff}(r)$, i.e., near $ r=0$.  In the
present approximation-theory context, the {\em practical}
irrelevance of the corresponding error terms is an interesting
consequence of the fact that {\em all} our approximate wave
functions  are exponentially decaying outside the
effective-potential valley at the sufficiently large $\ell$. This
makes the irregularity of the wave functions at $r=0 \ll T$ (which
is definitely there of course, together with the equally puzzling
``allowed tunnelling" to $r < 0$ \cite{Bjerrumh}) entirely
irrelevant. Still, it is worth noting that all these ``omissions"
definitely cause the ultimate divergence of the {\em infinite}
$1/\ell$ expansions, in the manner discussed thoroughly in the
specialized literature (of which we may recommend the most succinct
study \cite{Bjerrumd} to the interested reader's attention).


\newpage

\section*{References}

\begin{thebibliography}{00}

\bibitem{earth}
Moffatt H K 1978  {\it Magnetic field generation in electrically
conducting fluids} (Cambridge: University Press)

\bibitem{whereverderived}
Krause F and R\"adler K-H 1980 {\it Mean-field
magnetohydrodynamics and dynamo theory} (Berlin: Akademie-Verlag,
Oxford: Pergamon Press)

\bibitem{mhd-book-3}
Zeldovich Ya B, Ruzmaikin A A and Sokoloff D D 1983 {\it Magnetic
fields in astrophysics} (New York: Gordon \& Breach Science
Publishers)

\bibitem{ruediger-book} R\"udiger G and Hollerbach R 2004 {\it The magnetic
Universe} (Weinheim: Wiley-VCH)

\bibitem{high-conductivity}
Proctor M R E 1977 Astron. Nachr. {\bf 298} 19 and Geophys.
Astrophys. Fluid Dyn. {\bf 8} 311

\bibitem{high-conductivityb}
R\"adler K-H 1982 Geophys. Astrophys. Fluid Dyn. {\bf 20} 191

\bibitem{sokoloff} Sokoloff D D, Ruzmaikin A A and Shukurov A M 1983
Geophys. Astrophys. Fluid Dyn. {\bf 25} 293

\bibitem{Krause-Meinel} Krause F and Meinel R 1988 Geophys. Astrophys. Fluid Dyn. {\bf 43}
95

\bibitem{bar-shu-1} Baryshnikova Y and Shukurov A 1987 Astron. Nachr.
{\bf 308} 89
\bibitem{meinel} Meinel R 1989 Astron. Nachr. {\bf 310} 1

\bibitem{meinel-brandenburg} Meinel R and Brandenburg A 1990 Astron.
Astrophys. {\bf 238} 369

\bibitem{high-conductivityc}
K.-H. R\"adler K-H and U. Geppert U 1999
ASP Conference Series {\bf 178} 151




\bibitem{SG-prl} Stefani F and Gerbeth G 2005 Phys. Rev. Lett. {\bf
94}  184506; physics/0411050.

\bibitem{giesecke-2} Giesecke A, R\"udiger G and Elstner D 2005
Astron. Nachr. {\bf 326} 693, astro-ph/0509286

\bibitem{flips}
Stefani F, Gerbeth G, G\"unther U and Xu M 2006 Earth Planet. Sci.
Lett. {\bf 243} 828-840, physics/0509118

\bibitem{dynamoRiga}
Gailitis A, Lielausis O, Platacis E, Gerbeth G and Stefani F 2002
Rev. Mod. Phys. {\bf 74} 973

\bibitem{GS-jmp2003}
G\"{u}nther U and Stefani F 2003 J. Math. Phys. {\bf 44} 3097,
math-ph/0208012

\bibitem{jmpus}
G\"unther U,  Stefani F and Znojil M 2005
J. Math. Phys. {\bf 46} 063504, math-ph/0501069

\bibitem{GSG-cz2}
G\"{u}nther U, Stefani F and Gerbeth G 2004 Czech. J. Phys. {\bf
54} 1075, math-ph/0407015


\bibitem{GS-cz3}
G\"{u}nther U and Stefani F 2005  Czech. J. Phys. {\bf 55} 1099,
math-ph/0506021

\bibitem{Oleg}
G\"{u}nther U and Kirillov O N 2006
J. Phys. A: Math. Gen. {\bf 39} 10057, math-ph/0602013


\bibitem{cc}
Znojil M 2006
J. Phys. A: Math. Gen. {\bf 39} 441, quant-ph/0511085
and
 4047, quant-ph/0511194

\bibitem{Bjerrum}
Mlodinov L and Papanicolau N 1980 Ann. Phys. (N.Y.) {\bf 128} 314
and 1981 Ann. Phys. (N.Y.) {\bf 131} 1

\bibitem{Bjerrumb}
Sukhatme U and Imbo T 1983 Phys. Rev. {\bf D28}  418

\bibitem{Bjerrumc}
Mlodinov L and Shatz M P 1984 J. Math. Phys. {\bf 25} 943

\bibitem{Bjerrumd}
Bjerrum-Bohr N E J 2000
J. Math. Phys. {\bf 41}  2515, quant-ph/0302107

\bibitem{Bjerrume}
Fernandez F M 2001 {\it Introduction to Perturbation Theory in
Quantum Mechanics} (Boca Raton: CRC Press)

\bibitem{Bjerrumf}
Znojil M, Gemperle F and Mustafa O 2002 J. Phys. A: Math. Gen. {\bf
35} 5781, hep-th/0205181

\bibitem{Bjerrumg}
Mustafa O and Znojil M 2002
J. Phys. A: Math. Gen. {\bf 35}  8929, math-ph/0206042

\bibitem{Bjerrumh}
Znojil M 2003
J. Phys. A: Math. Gen. {\bf 36}
 9929, quant-ph/0307239

\bibitem{Bjerrumia}
B\'{\i}la H 2004 Czech. J. Phys. {\bf 54} 1049

\bibitem{Bjerrumi}
Znojil M 2004
Int. J. Pure Appl. Math. {\bf 12} 79, physics/0404123



\bibitem{Fluegge}
Landau L D and Lifshitz E M 1965 {\it Quantum mechanics:
non-relativistic theory} (Oxford: Pergamon Press)

\bibitem{onenpan} Baumg\"artel H 1984   {\it Analytic perturbation theory for
matrices and operators} (Berlin: Akademie-Verlag) and 1985 {\it
Operator Theory: Adv. Appl.  {\bf 15}} (Basel: Birkh\"auser)
%

\end{thebibliography}
 \end{document}